\documentclass[submission,copyright,creativecommons]{eptcs}
\providecommand{\event}{NCL'22} 
\usepackage{breakurl}             
\usepackage{underscore}           

\usepackage{longtable}
\usepackage{amssymb}
\usepackage{latexsym}
\usepackage{bussproofs}

\begin{document}

\title{Bilateral Inversion Principles}

\author{Nils K\" urbis
\institute{Department of Logic and Methodology of Science \\ University of \L\'od\'z \\ Poland}
\email{nils.kurbis@filhist.uni.lodz.pl}
}
\def\titlerunning{Bilateral Inversion Principles}
\def\authorrunning{N. K\" urbis}

\maketitle



\begin{abstract}
\noindent This paper formulates a bilateral account of harmony that is an alternative to one proposed by Francez. It builds on an account of harmony for unilateral logic proposed by K\"urbis and the observation that reading the rules for the connectives of bilateral logic bottom up gives the grounds and consequences of formulas with the opposite speech act. I formulate a process I call `inversion' which allows the determination of assertive elimination rules from assertive introduction rules, and rejective elimination rules from rejective introduction rules, and conversely. It corresponds to Francez's notion of vertical harmony. I also formulate a process I call `conversion', which allows the determination of rejective introduction rules from assertive elimination rules and conversely, and the determination of assertive introduction rules from rejective elimination rules and conversely. It corresponds to Francez's notion of horizontal harmony. The account has a number of features that distinguishes it from Francez's.
\end{abstract}

\section{Introduction} 
In Humberstone's \cite{humberstonerejection} and Rumfitt's \cite{rumfittyesno} bilateral formalisations of logic, the premises and conclusions of rules of inference are asserted and denied formulas. All formulas are \emph{signed} by $+$, indicating assertion, or $-$, indicating denial. Connectives are typically governed by four rules: those specifying the conditions under which a formula with the connective as main operator may be derived as asserted; those under which such formulas may be derived as denied; those that specify what follows from an asserted formula with the connective as main operator, possibly together with other premises or side deductions; and finally those that specify what follows when such formulas are denied, also possibly together with other premises or side deductions. The former are the assertive introduction and elimination rules of the connective, the latter its rejective introduction and elimination rules.\footnote{Exceptions to the four rule rule are $\bot$ and $\top$, the first of which can always be denied, never asserted, the latter always asserted, never denied, so each only has two rules. Arguably, however, the four rule rule may be salvaged by insisting that $\bot$ has the empty assertive introduction rule and $\top$ the empty rejective introduction rule, carrying over a point made by Prawitz from the unilateral to the bilateral case \cite[35]{prawitzmeaningandcompleteness}.} The major premise of an elimination rule is the one that contains the connective governed by the rule in the place in which it occurs in its schematic representation; as rules will be formulated in the present paper, this is always the leftmost formula. The minor premises are all the others, if any there are. Side-deductions are derivations of minor premises in which the rule permits the discharge of assumptions. Throughout the paper I shall employ Rumfitt's favourite deductive framework, single conclusion natural deduction, and free use will be made of certain structural rules. This is sufficient for present purposes, namely to formulate a bilateral counterpart to what Dummett calls \emph{harmony} \cite[215ff]{dummettLBM}, in turn closely related to what Prawitz calls \emph{the inversion principle} \cite[Ch II]{prawitznaturaldeduction}, that is suitable to Rumfitt's framework. Although harmony and inversion are easily transposed to an account that relates the assertive introduction and elimination rules and the rejective introduction and elimination rules for a connective, a question arises how to relate the assertive and the rejective rules. My purpose is the following: while Dummett's and Prawitz's notion of harmony is applicable to unilateral logic, Rumfitt lacks a comparable notion applicable to bilateral logic, and my aim is to provide one.\footnote{Notions of harmony suitable to other frameworks and other logics must be relegated to further work. But see footnote \ref{non-classical}.}

One such account has already been formulated by Francez. In response to a problem for bilateral proof-theoretic semantics posed by Gabbay \cite{gabbaybilateralism},\footnote{A similar problem in the form of bilateral versions of Prior's \emph{tonk} is posed in \cite{kurbisnormalbilateral}. The present paper provides a solution.} Francez \cite{francezgabbayresponse} slightly reformulates an earlier account \cite{francezbilateralharmony} of bilateral harmony. Francez formulates two inversion principles, or two notions of harmony. Extending a famous remark of Gentzen's -- that it is the introduction rules for a connective that define its meaning while its elimination rules are, in some sense, consequences thereof \cite[189]{gentzenuntersuchungen} -- to the bilateral case, Francez takes the assertive introduction rules for a connective to define its meaning (\cite[251]{francezbilateralharmony}, \cite[1656]{francezgabbayresponse}), and thus, following Dummett \cite[Ch 11]{dummettLBM}, to be self-justifying. What Francez calls \emph{vertical harmony} relates the assertive introduction rules for a connective to its assertive elimination rules, and its rejective introduction rules to its rejective elimination rules (\cite[250]{francezbilateralharmony}, \cite[1655]{francezgabbayresponse}): it allows the determination of the assertive elimination rules for a connective relative to its assertive introduction rules, and the determination of its rejective elimination rules relative to its rejective introduction rules. What Francez calls \emph{horizontal harmony} relates the assertive introduction rules for a connective to its rejective introduction rules (\cite[252ff]{francezbilateralharmony}, \cite[1656ff]{francezgabbayresponse}): it allows the determination of the rejective introduction rules for a connective from its assertive introduction rules. 

In this paper, I will offer an alternative account of bilateral harmony building on an account of unilateral harmony \cite[Ch 2.8]{kurbisproofandfalsity}\footnote{\label{non-classical}This notion of harmony was also proposed in \cite{kurbisstableharmony}. The account receives a very general treatment to cover a large range of sub-structural logics in \cite{nilsharmony}. It should be possible to adapt this and the present work to bilateral formulations of such logics.}  and the process of conversion \cite[Ch 7.4]{kurbisproofandfalsity} formulated for the rules of Wansing's bi-intuitionistic logic \cite{wansingmoregeneral}. In the next section I will note a difficulty with Francez's account. It is not insurmountable, but the fact that as it stands it cannot be quite correct is reason enough to pursue an alternative. The problem does not arise in the account presented here, which is also in some respects simpler than Francez's. My account has the further advantage of incorporating the following phenomenon. Proponents of proof-theoretic semantics usually opt for one kind of rules as those that define the meanings of the logical constants, the other kind being determined from them by an inversion principle or harmony: verificationists pick the introduction rules, pragmatists the elimination rules \cite[Ch 13]{dummettLBM}. If a choice must be made, however, it is plausible to permit that for some connectives, it is the introduction rules that define their meanings, while for others, it is their elimination rules that do so. For instance, while it may be more natural to take the introduction rules for $\land$ and $\supset$ as defining the meanings of these connectives, I should say that it is more natural to accord this property to the elimination rule in case of $\lor$. In the present account, every connective will be governed by rules of both these kinds: every connective is governed by one pair of introduction and elimination rules where it may be more natural to say that it is the introduction rule that define its meaning, and one pair in which this role is more naturally accorded to the elimination rule. This singular is correct: there will be one introduction rule, and one elimination rule. One pair are the assertive rules, the other the rejective rules for the connective. Moreover, informal elucidations to be given along the way of the present enquiry will contribute to conceptual clarification.\footnote{The account also rules out the  bilateral intuitionistic logic proposed by the author \cite{kurbisrumfitt}: its rules are not harmonious according to it. I argued that the system satisfies all the requirements Rumfitt imposes on a satisfactory bilateral logic; and so it does. Rumfitt did not have a precise enough notion of bilateral harmony at his disposal. This does not mean that there may not be other reasons why bilateralism, or some version thereof, coheres better with intuitionistic than with classical logic. See \cite{kurbisbilateraldetours} for an argument that Price's bilateralism \cite{pricesense,pricewhynot,pricenotagain} may be better served by intuitionistic logic.}

\section{A Problem with Francez's Vertical Harmony} 
It would take up too much space to go into the details of Francez's intricate account of bilateral harmony here. As it is easily available, there is also no need to do so, and I shall confine myself to giving a reason for choosing an alternative path.

Francez requires all elimination rules to be what von Plato calls \emph{general elimination rules} \cite{vonPlatoGenElim},\footnote{This form of elimination rules is also favoured by Read \cite{readGEharmony}.} that is, they are of the type of disjunction elimination and require side-deductions of minor premises that also provide the conclusion of the rule and above which assumptions are discharged. Francez's vertical harmony, however, does not permit the determination of a general elimination rule for a connective from its introduction rule, nor, indeed, of any other kind of elimination rule. Take one half of Francez's bilateral vertical inversion principle, that for the assertive rules governing a connective: 

\begin{quote}
Any conclusion of an assertion of $A\ast B$ is also a conclusion of any grounds for the assertion of $A\ast B$ \cite[250]{francezbilateralharmony}, where the grounds for an assertion of $A\ast B$ is the set of all coherent sets of assumptions of derivations of $+ \ A\ast B$ that end with $+ \ast I$ \cite[248]{francezbilateralharmony}, and a set of assumptions $\Gamma$ is coherent iff, for any $A$, not both, $\Gamma \vdash + \ A$ and $\Gamma\vdash - \ A$.
\end{quote}

\noindent Consider, for simplicity, conjunction. As I will only consider the assertive case, there is no need to sign formulas by $+$. Take its introduction rule as given: $A, B\vdash A\land B$. Suppose $C$ can be derived from $A\land B$. Then Francez's principle says that $C$ should also follow from $A$ and $B$. But that does not seem to justify the general elimination rules for conjunction, which is: 

\begin{prooftree} 
\AxiomC{$A\land B$}
\AxiomC{$\underbrace{[A]^i, [B]^i}$}
\noLine
\UnaryInfC{$\Pi$}
\noLine
\UnaryInfC{$C$}
\RightLabel{$_i$}
\BinaryInfC{$C$}
\end{prooftree} 

\noindent This rule specifies what follows from $A\land B$ given what follows from $A$ and $B$ and discharging those latter assumptions. Francez's inversion principle does something else, namely to specify what follows from $A$ and $B$ given what follows from $A\land B$. Now presumably this should no longer depend on the assumption $A\land B$, so that that formula is discharged. Thus, rather than justifying the general elimination rule for $\land$, Francez's inversion principle justifies the following rule: 

\begin{prooftree}
\AxiomC{$A$}
\AxiomC{$B$}
\AxiomC{$[A\land B]^i$}
\noLine
\UnaryInfC{$\Pi$}
\noLine
\UnaryInfC{$C$}
\RightLabel{$_i$}
\TrinaryInfC{$C$}
\end{prooftree} 

\noindent This, however, is no elimination rule for $\land$ at all, but rather  (under assumptions fulfilled by at least classical, intuitionistic and minimal logic) a rule interderivable with conjunction introduction. In fact, it is what Negri and von Plato call the \emph{general introduction rule} for conjunction \cite[217]{negriplatostructural}.\footnote{For more on general introduction rules, and arguments why it is they that really capture the essence of what it is to be an introduction rule, see \cite{milneinversion}. Milne proposes a system of classical logic with general introduction and elimination rules which has the remarkable property that for any deduction in it, there is one with the subformula property \cite{milnesubformula}. A constructive proof of this result as a consequences of a normalisation theorem for Milne's system is in \cite{kurbisgenrulescl}. Similar results for Negri and von Plato's intuitionistic system with general introduction and elimination rules are proved in \cite{kurbisgenrulesintuitionist}.} The motivation of Francez's account of bilateral harmony is thus flawed, and it stands in need of conceptual clarification. The present paper contributes to the latter, although the resulting account will differ from Francez's. The alternative to be presented does, however, I think, capture the spirit, if not the letter, of Francez's account.

\section{Rules Top Down and Bottom Up}
The sequents of a sequent calculus can be read in two ways. $\Sigma \Rightarrow \Delta$ is usually read from left to right, its semantic interpretation being that if all formulas of $\Sigma$ are true, some formula of $\Delta$ is true. We can, however, also read it from right to left, its semantic interpretation then being that if all formulas in $\Delta$ are false, then some formula in $\Sigma$ is false. Something similar, if not quite so neatly symmetric, is the case for rules of inference of systems of natural deduction. They, too, can be read in two ways. Usually they are read from top down as preserving truth: somewhat simplified, if the premises of the rule are true, then the conclusion is also true. But we can also read them bottom up: if the conclusion is false, at least one of the premises is false. Then the rules preserve falsehood. As an example, consider conjunction:  

\begin{center}
\AxiomC{$A$} 
\AxiomC{$B$}
\LeftLabel{$\land I$: \ }
\BinaryInfC{$A\land B$}
\DisplayProof\qquad\qquad
\AxiomC{$A\land B$}
\LeftLabel{$\land E$: \ }
\UnaryInfC{$A$}
\DisplayProof\quad 
\AxiomC{$A\land B$} 
\UnaryInfC{$B$}
\DisplayProof
\end{center}

\noindent Read top down, the rules confer truth from the premises to the conclusions: by $\land I$, if $A$ and $B$ are truth, then $A\land B$ is true, and by $\land E$, if $A\land B$ is true, then $A$ is true and $B$ is also true. Read bottom up, the rules confer falsity from the conclusions to at least one of the premises: by $\land E$, if $A$ is false, so is $A\land B$, and the same if $B$ is false, and by $\land I$, if $A\land B$ is false, at least one of $A$ and $B$ is false. 

The disjunctive conclusion to which we are led if we read $\land I$ bottom up is not congenial to framework of natural deduction chosen here. However, the situation is no different from the situation in disjunction elimination.Thus, by $\land I$, if $A\land B$ is false, and the same conclusion $C$ follows from the falsity of $A$ and also from the falsity of $B$, then $C$ follows from the falsity of $A\land B$. At this point the earlier simplification in reading rules top down as truth preserving becomes apparent: account needs to be taken of the discharge of assumptions, in this case assumptions of the falsity of $A$ and of the falsity of $B$. 

According to (the verificationist version of) proof-theoretic semantics, the introduction rules for a connective $\ast$ define its meaning by laying down the canonical grounds for asserting a sentence with the connective as main operator. Reading the introduction rules top down, that is. But of course the very same introduction rules also give information in case there are grounds for refraining from asserting $A\ast B$: then some of the grounds for asserting $A\ast B$ as specified in the introduction rule do not obtain. 

To apply these observations to the bilateral case, we need some terminology. Lower case Greek letters range over signed formulas. $\alpha^\ast$ designates the \emph{conjugate} of $\alpha$, the result of reversing its sign from $+$ to $-$ and conversely. The primitive connectives of Rumfitt's formalisation of classical bilateral logic are $\supset$, $\land$, $\lor$ and $\neg$, and each of them is governed by four rules, its assertive and rejective introduction and elimination rules \cite[800ff]{rumfittyesno}. There is also a rule for the \emph{co-ordination} of assertion and denial or a \emph{Co-ordination Principle}: if $\alpha$ and $\alpha^\ast$ follow from $\Gamma, \beta$, then $\beta^\ast$ follows from $\Gamma$.\footnote{For a normalisation theorem for Rumfitt's system, see \cite{kurbisnormalbilateral}. \cite{kurbisnormalbilateralnote} corrects an error in one of the reduction steps.} The co-ordination principle formalises a condition on the coherence of assertion and denial.\footnote{Alternatively, there are rules governing an expression $\bot$: $\alpha, \alpha^\ast\vdash\bot$, and if $\Gamma, \alpha\vdash\bot$, then $\Gamma\vdash\alpha^\ast$. $\bot$ then marks the incompatibility between the speech acts assertion and denial. As I am using $\bot$ for another purpose, however, namely as a proposition that is always false, it is preferable not to use these rules here. Nothing essential hangs on this.}

Again simplifying somewhat, according to bilateralists, the rules of logic preserve assertibility and deniability. As simple examples, consider the assertive rules for conjunction and the rejective rules for disjunction of Rumfitt's formalisation of classical bilateral logic. Read top down as intended, the assertive rules for conjunction transmit assertibility from the premises to the conclusions of the rules, and the rejective rules for disjunction transmit deniability: 

\begin{center}
\AxiomC{$+ \ A$}
\AxiomC{$+ \ B$}
\LeftLabel{$+ \land I$: \ }
\BinaryInfC{$+ \ A\land B$}
\DisplayProof\qquad\qquad
\AxiomC{$+ \ A\land B$}
\LeftLabel{$+\land E$: \ }
\UnaryInfC{$+ \ A$}
\DisplayProof\quad 
\AxiomC{$+ \ A\land B$}
\UnaryInfC{$+ \ B$}
\DisplayProof

\bigskip

\AxiomC{$- \ A$}
\AxiomC{$- \ B$}
\LeftLabel{$-\lor I$: \ }
\BinaryInfC{$- \ A\lor B$}
\DisplayProof\qquad\qquad
\AxiomC{$- \ A\lor B$}
\LeftLabel{$-\lor E$:  \ }
\UnaryInfC{$- \ A$}
\DisplayProof\quad
\AxiomC{$- \ A\lor B$}
\UnaryInfC{$- \ B$}
\DisplayProof
\end{center} 

\noindent Just as unilateral rules give information not only about what is the case if their premises are true, but also if their conclusions are false, bilateral rules also give information about what is the case if the opposite speech act is applied to their conclusions. If $A$ is denied and $A\land B$ were asserted, then by $+\land E$ there would be an incoherence, i.e. that between $A$ being denied and asserted, and hence $A\land B$ must be denied. Similarly if $B$ is denied. If $B$ is asserted and $A\lor B$ were denied, then by $-\lor E$ there would be an incoherence, i.e. that between $B$ being asserted and denied, and hence $A\lor B$ must be asserted. Similarly if $A$ is asserted. Thus read bottom up, the two rules $+\land E$ give Rumfitt's introduction rules for denied conjunctions and the two rules $-\lor E$ give Rumfitt's introduction rules for asserted disjunctions:

\begin{center}
\AxiomC{$- \ A$}
\LeftLabel{$-\land I$: \ }
\UnaryInfC{$- \ A\land B$}
\DisplayProof\quad
\AxiomC{$- \ B$}
\UnaryInfC{$- \ A\land B$}
\DisplayProof\qquad\qquad
\AxiomC{$+ \ A$}
\LeftLabel{$+\lor I$:  \ }
\UnaryInfC{$+ \ A\lor B$}
\DisplayProof\quad
\AxiomC{$+ \ B$}
\UnaryInfC{$+ \ A\lor B$}
\DisplayProof
\end{center}

\noindent Clearly, we could have started with these rules, too, and, reading them bottom up, determined the assertive elimination rules for $\land$ and the rejective elimination rules for $\lor$. 

Generalising this observation, in the simplest cases reading the rules of bilateral logic bottom up tells us how to use the formulas occurring in those rules with their signs reversed, i.e. they specify how to use the conjugates of those formulas in deductions. 

Reading $+\land I$ and $-\lor I$ bottom up leads to the disjunctive case already noted in the unilateral case. If $A\land B$ is denied, then either $A$ or $B$ must be denied. We may not know which one, but if the same signed formula follows from both of them, it must be correct independently of which of $A$ and $B$ is to be denied. Thus whichever it may be, if the same signed formula follows from both of them, it follows from $-\ A\land B$. Similarly, if $A\lor B$ is asserted: then either $A$ or $B$ must be asserted, but whichever it may be, if the same signed formula follows from both of them, it follows from $+\ A\lor B$. The results are Rumfitt's elimination rules for denied conjunctions and asserted disjunctions, where $\phi$ ranges over signed formulas: 

\begin{center}
\AxiomC{$- \ A\land B$}
\AxiomC{$[- \ A]^i$}
\noLine
\UnaryInfC{$\Pi$}
\noLine
\UnaryInfC{$\phi$}
\AxiomC{$[- \ B]^i$}
\noLine
\UnaryInfC{$\Sigma$}
\noLine
\UnaryInfC{$\phi$}
\LeftLabel{$-\land E$: \ }
\RightLabel{$_i$}
\TrinaryInfC{$\phi$}
\DisplayProof\qquad\qquad
\AxiomC{$+ \ A\lor B$}
\AxiomC{$[+ \ A]^i$}
\noLine
\UnaryInfC{$\Pi$}
\noLine
\UnaryInfC{$\phi$}
\AxiomC{$[+ \ B]^i$}
\noLine
\UnaryInfC{$\Sigma$}
\noLine
\UnaryInfC{$\phi$}
\LeftLabel{$+\lor E$: \ }
\RightLabel{$_i$}
\TrinaryInfC{$\phi$}
\DisplayProof
\end{center} 

\noindent These rules exhibit that it was too simple to say that bilateral logical rules preserve assertibility and deniability: $\phi$ can be either speech act. In constructing a deduction, however, having reached $- \ A\land B$, one proceeds to assuming $- \ A$ and $- \ B$ with the aim of deriving the same signed formula from each; analogously for disjunction. So one does proceed to formulas of the same sign as the major premise, only not as a conclusion but as assumptions to be discharged. It would, however, still be too simplistic to say that assertibility and deniability are transmitted from premises to conclusions or from premises to assumptions to be discharged further down the line, as will become clear when the rules for implication and negation are given. None of this affects what is at issue: the notion of preservation of assertibility and deniability was introduced for heuristic purposes and can be left behind once the general method of reading rules bottom up has been exhibited. I shan't have occasion to come back to the notion.\footnote{Not that there may not be an interesting issue here: according to the unilateralist, rules of logic preserve truth. The bilateralist owes us an answer to the question what it is that bilateral rules preserve.}

More importantly, therefore, we have reconstructed the four rules for each connective. The rejective rules for conjunction can be determined from their assertive rules and the assertive rules for disjunction can be determined from their rejective rules. These are the roots for an account of bilateral inversion. 

The result is roughly the following: 

\begin{quote}
\noindent \emph{Bilateral Harmony I.} If $G$ are the grounds for asserting a formula $A \ast B$ (as specified by an assertive introduction rule for $\ast$), then denying $A\ast B$ gives grounds for denying that some of $G$ obtain. 

\noindent \emph{Bilateral Harmony II.} If $G$ are the consequences of asserting $A\ast B$ (as specified by an assertive elimination rule for $\ast$), then denying that some of $G$ obtain gives grounds for denying $A\ast B$.  
\end{quote}

\noindent This is not perfectly precise and does not obviously apply to all the rules for all the connectives, but the discussion to follow will contribute to its clarification. 

We can also try to read $+\lor E$ and $-\land E$ bottom up, but how to do so is less perspicuous than in the cases discussed. Notice, however, that for a satisfactory account of bilateral harmony it is not required that all rules can be read bottom up. It suffices that we can determine all three other rules for a connective from any given one, and, as we will see, this can be achieved by reading only \emph{some} rules bottom up and using a different method for the rest. 

To give a complete account of bilateral harmony, I propose to look at one for unilateral harmony first, which is then extended to bilateral logic by combining it with the method of reading rules bottom up.

\section{Unilateral Harmony} 
The intuitive idea behind the demand for harmony between the introduction and elimination rules for a connective $\ast$ is that its elimination rules should \emph{not} allow the derivation of \emph{more} conclusions from $A\ast B$ than one is entitled to draw given the grounds for the assertion of $A\ast B$ as specified by the introduction rules for $\ast$, and furthermore that its elimination rules should allow the derivation of \emph{all} conclusions one is entitled to draw from $A\ast B$.\footnote{\label{stability}Strictly speaking, this is the requirement Dummett calls \emph{stability} \cite[287]{dummettLBM}; harmony concerns only the `no more' part of the above. It is a fairly established use of terminology to speak only of harmony, which I here follow, although stability is meant.} For the introduction and elimination rules for a connective to define its meaning completely, proof-theoretic semantics demands that there be a perfect balance between introduction and elimination rules: The grounds for deriving $A\ast B$ as specified by $\ast I$ determine the consequences of $A\ast B$, i.e. $\ast E$, and conversely, the consequences of $A\ast B$ as specified by $\ast E$ determine the grounds of $A\ast B$, i.e. $\ast I$. If the rules for a connective are in harmony, then given $\ast I$, we should be in a position to determine $\ast E$, and conversely, given $\ast E$ we should be in a position to determine $\ast I$. Accordingly, in this section I will specify a method for ``reading off'' the elimination or introduction rules for a connective from its introduction or elimination rules.\footnote{The account is presented in detail in \cite[Ch 2]{kurbisproofandfalsity}, to which I refer the interested reader for further discussion, in particular of the philosophical motivation of the account. For an alternative account to justify logical rules focussing on the notions of conservative extension and molecularity in the theory of meaning,  but with a close connection to harmony, see \cite{weisskurbismolecularity}.}

More precisely, I will give two such methods, corresponding to the two kinds of rules one finds in formalisations of intuitionistic logic in a standard system of natural deduction. There is a difference in the rules for conjunction and those for disjunction which is seen in their elimination rules: the elimination rule for disjunction requires side deductions of an arbitrary formula, which is also the conclusion of the rule, from assumptions discharged by the rule; the elimination rules for conjunction do not require side deductions. I will call `rules of type 1' those of that are of the type of conjunction and `rules of type 2' those that are of the type of disjunction. $\supset$ is governed by rules of type 1. $\top$ is a limiting case of a connective governed by rules of type 1. $\bot$ is a limiting case of a connective governed by rules of type 2. Whether an introduction rule is of type 1 or of type 2 cannot always be determined by looking at the rule alone: in some cases it is only the pair of introduction and elimination rules that exhibit their type. 

Rules of inference are harmonious if and only if they conform to one of the following two types: 

\begin{quotation}
\noindent \emph{Rules of Type 1:} 

\noindent A connective $\ast$ is governed by rules of type 1 if it has exactly one introduction rule, which can be of any form whatsoever, as long as the conclusion of the rule is constructed by connecting all and only the premises and discharged hypotheses of the rule by using $\ast$ as main operator. 

The elimination rules for $\ast$ are determined in this way: To each premise of the introduction rule there corresponds an elimination rule which has that premise as its conclusion and if there are discharged hypotheses above it, these become minor premises of the elimination rule. The major premise of the elimination rule is the conclusion of the introduction rule.
\end{quotation} 

\noindent It is not difficult to see that the elimination rules for $\land$ and $\supset$ of intuitionistic logic can be determined from their introduction rules by this method. A special case is $\top$: its introduction rule $\overline{\ \top \ }$ has no premises, and therefore it has no elimination rule. I leave it to the reader to give the converse of this process which determines the introduction rule from the elimination rules of a connective governed by rules of type 1. 

\begin{quotation}
\noindent \emph{Rules of Type 2:} 

\noindent A connective $\#$ is governed by rules of type 2 if it has exactly one elimination rule, which can be of any form whatsoever as long as its major premise is constructed using $\#$ as main operator from all and only the discharged hypotheses of collateral deductions of minor premises of formulas that are the same as the conclusion of the rule. 

The introduction rules for $\#$ are determined in this way: To each collateral deduction of the elimination rule there corresponds an introduction rule which has as its premises the discharged hypotheses of that collateral deduction. The conclusion of the introduction rule is the major premise of the elimination rule.
\end{quotation} 

\noindent It is not difficult to see that the introduction rules for $\lor$ of intuitionistic logic can be determined from its elimination rule by this method. A special case is $\top$: its elimination rule has no minor premises, and hence it has no introduction rule. I leave it to the reader to give the converse of this process which determines the elimination rule from the introduction rules of a connective governed by rules of type 2. 

In the following I will call the process just described `inversion'. For further discussion and examples, the reader may wish to consult \cite[Ch 2.8]{kurbisproofandfalsity}. The connectives of intuitionist logic are all harmonious according to this definition of harmony. 

For an example of rules that are not harmonious, consider Prior's \emph{tonk} \cite{priorrunabout}: 

\begin{center}
\AxiomC{$A$}
\UnaryInfC{$A tonk B$}
\DisplayProof\qquad\qquad
\AxiomC{$A tonk B$}
\UnaryInfC{$B$}
\DisplayProof
\end{center}

\noindent Its rules do not conform to one of the two types. Its introduction rule is not of type 1, but its elimination rule is not of type 2. Hence they are not harmonious. The rules governing a connective define its meaning only if they are in harmony. Hence the rules governing $tonk$ do not define its meaning. As nothing else has been given to do so, we must conclude that $tonk$ is meaningless. 

To close this section, let's look again at the debate between verificationists and pragmatists in proof-theoretic semantics. According to verificationists, the introduction rules for a connective define its meaning, while its elimination rules are determined from them by an inversion principle or harmony; according to pragmatists it is the other way round and the elimination rules define meanings, with the introduction rules determined from them. Both, I should say, have a point, but not necessarily in the generality with which it is put forward. We may draw on the insights of both parties and accord a certain priority to the introduction rules of some connectives, while for others we chose the elimination rules as prior. The meanings of some connectives are defined more naturally by their introduction rules, while for others this role is taken by the elimination rules. With some connectives we look at the grounds to define their meanings, with others to the consequences. It is plausible that the meaning of $\lor$ is defined in terms of $\lor E$ rather than in terms of the the two rules of $\lor I$. (Why should $A\lor B$ follow for arbitrary $B$ from $A$ and for arbitrary $A$ from $B$? Answer: because these rules are harmonious relative to $\lor E$, and a similar question does not arise for proof by cases.) For implication, I should say it is the introduction rule that is more important, as $A\supset B$ is supposed to express that $B$ may be inferred from $A$. For $\top$, we must chose the introduction rule, as it has no elimination rule, for $\bot$ it is the other way round. Conjunction may be an exception, and maybe here it really does not matter which rules to chose, those of type 1 or the general elimination rules of type 2, which at least in classical, intuitionist and minimal logic are interderivable. 

Generalising, I should say that for rules of type 1, it is more natural to begin with an introduction rule and to give a method for ``reading off'' corresponding elimination rules, while for rules of type 2, it is more natural to begin with an elimination rule and to give a method for ``reading off'' corresponding introduction rules, as I have done above. Thus for rules of type 1, it is more natural to accord priority to the introduction rules, while for rules of type 2 it is more natural to chose the elimination rules. Following this line of thought, it would be natural to say that their introduction rule defines the meanings of connectives governed by rules of type 1, whereas it is its elimination rule that does so if a connective is governed by rules of type 2. 

It should be kept in mind, however, that it is not just the elimination rules or just the introduction rules that define the meanings of connectives. Rather, it is the introduction/elimination rules plus harmony, on the basis of which the correct elimination/introduction rules are determined. In a sense, then, it does not matter which rules are chosen as the ones defining meaning. And indeed, in each case, the process could be reversed, and an introduction rule specified relative to the elimination rules, in the case of rules of type 1, and an elimination rule specified relative to the introduction rules, in the case of rules of type 2. At this point it is worth being a little more precise with the terminology. According to Dummett, the rules governing a connective define its meaning completely if and only if these rules are stable, which is harmony plus its converse.\footnote{See footnote \ref{stability}. Other conditions may need to be in place, too, but these need not concern us here. See \cite[Ch 11-13]{dummettLBM}, also for the rest of what is said in this paragraph. For doubt concerning the left to right direction of the equivalence above, see \cite{kurbisnegation}.} Connectives that are not governed by stable rules require something other than purely logical rules in terms of which the grounds and consequences of assertions with them as main operator are specified and hence their meaning determined.\footnote{For instance, I have argued that the rules governing popular modal operators in standard systems of natural deduction, such as those of S4 given by Biermann and de Paiva \cite{biermannpaiva}, are governed by merely harmonious, but not stable, rules. See again \cite[Ch 2]{kurbisproofandfalsity} and also \cite{kurbissketch}. The latter contains a proposal of how to formulate stable rules for necessity, building on work by Pfenning and Davis \cite{pfenningdaviesmodal} and Do\v sen \cite{dosensequentmodal,dosenintmodal}.} In the case of stable rules, the two processes of determining elimination rules from introduction rules and introduction rules from elimination rules lead to the same result (or at least to interderivable rules). Pragmatists and verificationists agree on the form of stable rules of inference, that is, they agree on the meanings of the connectives governed by such rules. Thus in these cases, there is a clear sense in which it does not matter whether we chose the verificationist or the pragmatist path. This does not affect, of course, what has been said previously: even this being so, there may be a more natural choice depending on the connective under consideration. 

This discussion will undoubtedly have left some readers dissatisfied, but for more detailed discussion I must refer the reader to the literature quoted and further development must await another occasion.

\section{Bilateral Harmony} 
The classification of rules of inference into those of type 1 and those of type 2 can clearly be extended to bilateral systems: all one needs to do is to add the signs $+$ and $-$ indicating asserted and denied formulas. So adjusted, inversion works for bilateral logic: the signs of formulas are carried over from introduction to elimination rules or conversely, the minor premises and conclusions of elimination rules of type 2 are changed into an expression ranging over signed formulas. Inversion can be adapted to determine assertive elimination from assertive introduction rules and conversely, and rejective elimination from rejective introduction rules and conversely. It corresponds to Francez's vertical harmony. What is needed now is a process that corresponds to Francez's horizontal harmony and allows the determination of rejective rules for a logical constant from its assertive rules, or of assertive rules from its rejective rules. It suffices that some assertive rules are determined from some rejective rules or some rejective rules from assertive rules, as long as inversion does the rest. A satisfactory account of bilateral harmony must let us determine the remaining three sets of rules for a logical constant from any given one, and inversion adjusted to the formalism of bilateral logic achieves one half of this already. 

In Rumfitt's system conjunction and disjunction are governed by rules of type 1 as well as by rules of type 2. Conjunction has assertive rules of type 1 and rejective rules of type 2. Disjunction has rejective rules of type 1 and assertive rules of type 2. The suggestion presents itself that each connective be governed by both types rules. Following up on the thoughts expressed in the previous section, this means that every connective has both, a verificationist and a pragmatist aspect to its meaning. For some connectives, the verificationist aspect resides in their rejective rules, the pragmatist aspect in their assertive rules; for others it is the other way round. For instance, the verificationist aspect of the meaning of disjunction resides in its rejective rules, its pragmatist aspect in its assertive rules; for conjunction it is the other way round. 

In Rumfitt's formulation of bilateral logic, however, implication has only got rules of type 1: its assertive rules are those of unilateral intuitionistic logic with $+$ prefixed to each formula; its rejective introduction rule is $+ \ A, - \ B\vdash - \ A\supset B$, its rejective elimination rules are $- \ A\supset B \vdash + \ A$ and $- \ A\supset B \vdash - \ B$. It is straightforward to find a rejective elimination rule of type 2 for implication. $+\!\supset\! I$ lays down that $A\supset B$ may be asserted if the assertion of $B$ can be deduced from the assertion that $A$, i.e. if $B$ is asserted if $A$ is. If $A\supset B$ is denied, this is not the case, i.e. it is not the case that $B$ is asserted if $A$ is, and so $A$ is asserted and $B$ denied.\footnote{At least given the notion of consequence in use here, but this is not at issue.} Hence anything that follows from $+ \ A$ and $- \ B$ follows from $- \ A\supset B$: 

\begin{prooftree} 
\AxiomC{$-\ A\supset B$}
\AxiomC{$\underbrace{[+ \ A]^i, [- \ B]^i}$}
\noLine
\UnaryInfC{$\Pi$}
\noLine
\UnaryInfC{$\phi$}
\LeftLabel{$-\!\supset\! E$: \ }
\RightLabel{$_i$}
\BinaryInfC{$\phi$}
\end{prooftree} 

\noindent Applying inversion to this rule yields Rumfitt's rejective introduction rule: 

\begin{prooftree}
\AxiomC{$+ \ A$}
\AxiomC{$- \ B$}
\LeftLabel{$-\!\supset\! I$: \ }
\BinaryInfC{$- \ A \supset B$}
\end{prooftree} 

\noindent Rumfitt's logic also has rules for a primitive negation operator, but no rules for the propositional constant $\bot$. I will come back to them later. 

I shall call the process of turning an assertive rule into a rejective rule or conversely \emph{conversion}. It corresponds to Francez's horizontal harmony. Conversion employs the observation that reading rules bottom up gives the consequences of formulas with the sign of the conclusion reversed. Reading an introduction rule bottom up yields an elimination rule. We could formulate the process of conversion as one that determines elimination rules of type 2 from introduction rules of type 1 and conversely. From the conceptual point of view, however, it is more natural to formulate conversion so as to determine introduction rules of type 2 from elimination rules of type 1 and conversely, because the process of reading rules bottom up is less evidently applicable to elimination rules of type 2 than to introduction rules of that type. As demonstrated in the cases of conjunction and disjunction, the determination of the grounds for denying $A\land B$ from the consequences of asserting it, and conversely, and the determination of the consequences of denying $A\lor B$ from the grounds for asserting it, and conversely, are more straightforward than the determination of the consequences of denying $A\land B$ from the grounds for asserting it, and conversely, and the determination of the consequences of asserting $A\lor B$ from the grounds for denying it, and conversely. The version of conversion to be presented now generalises this process.  

To make the formulation of the process of conversion more manageable, I impose the following restrictions on the introduction rules for the connectives: 

\begin{quote} 
1.i If an introduction rule of type 1 discharges more than one assumption, exactly one of them has the sign of the conclusion. 

2.i If an introduction rule of type 2 has more than one premise, exactly one of them has the same sign as the conclusion.\end{quote}

\noindent These restrictions guarantee that there is a principled way of deciding which premise is turned into the conclusion in conversion and ensure the uniqueness of the rules so determined. By inversion, they correspond to the following restrictions on elimination rules: 

\begin{quote}
1.e If an elimination rule of type 1 has more than one minor premise, exactly one of them has the sign of the major premise.

2.e If an elimination rule of type 2 discharges more than one assumption above a minor premise, exactly one of them has the sign of the major premise.
\end{quote} 

\noindent The restrictions are not strictly necessary. They are imposed for convenience to enable an easier illustration of the process of conversion than would be possible if full generality was achieved. Furthermore, in the presence of Non-Contradiction and Reductio, rules determined by a more general process of conversion would be interderivable. To establish this is a needless complication at this point. The aim of the present paper is to propose an alternative to Francez's account of bilateral harmony, not to propose its most general version. This is left for future occasions. 

Now on to conversion. There are two processes: 

\begin{quote}
1. \emph{To convert an elimination rule $\rho$ of type 1 of a signed formula into an introduction rule $\rho'$ of type 2 for its conjugate:} 

Turn the major premise of $\rho$ with its sign reversed into the conclusion of $\rho'$, and the conclusion of $\rho$ with its sign reversed into a premise of $\rho'$, and keep any other premises of $\rho$ as premise of $\rho'$. Do so for all the elimination rules of the connective.

\bigskip

2. \emph{To convert an introduction rule $\rho$ of type 2 of a signed formula into an elimination rule $\rho'$ of type 1 for its conjugate}: 

Turn the conclusion of $\rho$ with its sign reversed into the major premises of $\rho'$, and (i) if $\rho$ has more than one premise, turn the premise with the same sign as the conclusion with its sign reversed into the conclusion of $\rho'$, and keep the other premise of $\rho$ as a premise of $\rho'$, otherwise (ii) turn the premise into the conclusion with its sign reversed. Do so for all the introduction rules of the connective.
\end{quote}

\noindent Both process correspond to the informally motivated observation that applying the opposite speech act to the conclusion of an introduction rule for its main connective gives its consequences together with the minor premises as specified by the introduction rule.\footnote{The following process would work to turn an introduction rule of type 1 into an elimination rule of type 2 of the conjugate formula: the conjugate of the conclusion of the introduction rule is the major premise of the elimination rule, to each premise of the introduction rule corresponds a side deduction of an arbitrary signed formula $\phi$ above which the conjugate of the premise of the introduction rule and any signed formulas that are above it are discharged. Conversely, to turn an elimination rule of type 2 into an introduction rule of type 1, take each side deduction and form an introduction rule out of it by using the major premise as the conclusion and one of the discharged assumptions as a premise, the other as a discharged assumption above it. It is not quite so clear, however, what this process has to do with the conceptual considerations of reading off the grounds for deriving the conjugate formula from an elimination rule for the signed formula.} 
  
Every connective is required to be governed by assertive and rejective rules and by rules of both types. Together, inversion and conversion determine the three remaining rules for a connective from any given one of the four: 

\begin{quote}
\noindent Given elimination rules or an introduction rule of type 1, determine the harmonious introduction rule or elimination rules by inversion. Take each elimination rule and turn it into an introduction rule of type 2 by conversion. Determine the harmonious elimination rule by inversion. 

\noindent Given introduction rules or an elimination of type 2, determine the harmonious elimination rule or introduction rules by inversion. Take each introduction rule and turn it into an elimination rule of type 1 by conversion. Determine the harmonious introduction rule by inversion. 
\end{quote}

\noindent The procedure works for $\bot$ and $\top$. $\bot$ has the assertive elimination rule of type 2 $+ \ \bot \vdash + A$, and so it has no assertive introduction rule; thus by conversion, it has no rejective elimination rule, and so by inversion it has the rejective introduction rule $\vdash - \ \bot$. Analogously for $\top$, which has the assertive elimination rule of type 1 $\vdash + \ \top$.

To close the discussion, let's have a brief look at negation. In Rumfitt's system, it is governed by the four rules $- \ A\vdash + \ \neg A$, $+ \ \neg A\vdash - \ A$, $+ \ A\vdash - \ \neg A$ and $- \ \neg A\vdash + \ A$. In the present account, where $\bot$ is a propositional constant, there is no need for primitive negation rules, as negation may be defined in terms of $\supset$ and $\bot$. It is worth nonetheless to consider which rules for a primitive negation operator would fit the present proposal. Rumfitt's rules don't: there is no rule of type 2 governing negation. Fitting rules are forthcoming by looking at the rules for $\supset$ given earlier when the consequent of the implication is $\bot$. Then the rules of type 1 for $\neg$ are its assertive introduction and elimination rules, its rules of type two the rejective introduction and elimination rules: 

\begin{quote}
(i) From $- \ A$ together with $+ \ A$  derive $+ \ \bot$ by the co-ordination principle, then $+ \ \neg A$ by $+\!\supset\! I$, discharging $+ \ A$. 

(ii) From $+ \ \neg A$ together with $+ \ A$ derive $+ \ \bot$ by $+\!\supset\! E$, then $- \ A$ by the co-ordination principle, as $\vdash - \ \bot$, discharging $+ \ A$. 

(iii) From $+ \ A$ and $\vdash - \ \bot$ infer $- \ \neg A$ by $-\!\supset\! I$. 

(iv) Given a deduction of $\phi$ from $+ \ A$, by $-\!\supset\! E$ there is a deduction of $\phi$ from $-  \ \neg A$, discharging $+ \ A$ and vacuously discharging $- \ \bot$. 
\end{quote}

\noindent Thus we may use the following rules for a primitive negation operator: 

\begin{center}
\bottomAlignProof
\AxiomC{$- \ A$}
\LeftLabel{$+\neg I$:}
\UnaryInfC{$+ \ \neg A$}
\DisplayProof\quad
\bottomAlignProof
\AxiomC{$+ \ \neg A$}
\LeftLabel{$+\neg E$:}
\UnaryInfC{$- \ A$}
\DisplayProof\qquad
\bottomAlignProof
\AxiomC{$+ \ A$}
\LeftLabel{$-\neg I$:}
\UnaryInfC{$- \ \neg A$}
\DisplayProof\quad
\bottomAlignProof
\AxiomC{$- \ \neg A$}
\AxiomC{$[+ \ A]^i$}
\noLine
\UnaryInfC{$\Pi$}
\noLine
\UnaryInfC{$\phi$}
\RightLabel{$_i$}
\LeftLabel{$-\neg E$:}
\BinaryInfC{$\phi$}
\DisplayProof
\end{center} 

\noindent As these rules avoid appeal to $\bot$, they define the meaning of negation within the bilateral framework directly in terms of the conditions for the assertion and the denial of $\neg A$ and their consequences. 

The rules justified by the processes of inversion and conversion are those of classical bilateral logic. 

Finally, here are some examples of connectives which, although they satisfy inversion, do not satisfy conversion. Consider the connective \emph{conk}, governed by the following rules: 

\begin{center}
\AxiomC{$+ \ A$}
\AxiomC{$+ \ B$}
\LeftLabel{$+ \mathit{conk} I$:} 
\BinaryInfC{$+ \ A \mathit{conk} B$}
\DisplayProof \qquad
 \AxiomC{$+ \ A\mathit{conk} B$}
\LeftLabel{$+\mathit{conk} E$:}
\UnaryInfC{$+ \ A$}
\DisplayProof \quad 
\AxiomC{$+ \ A\mathit{conk} B$}
\UnaryInfC{$+ \ B$}
\DisplayProof

\bigskip

\AxiomC{$- \ A$}
\AxiomC{$- \ B$}
\LeftLabel{$-\mathit{conk}I:$}
\BinaryInfC{$- \ A\mathit{conk} B$}
\DisplayProof\qquad
\AxiomC{$- \ A\mathit{conk} B$}
\LeftLabel{$-\mathit{conk} E$:}
\UnaryInfC{$- \ A$}
\DisplayProof\quad
\AxiomC{$- \ A\mathit{conk} B$}
\UnaryInfC{$- \ B$}
\DisplayProof
\end{center} 

\noindent If $conk$ is present in the logic, then, given the co-ordination principle, the assertion of any formula follows from the assertion of any formula, and the denial of any formula follows from the denial of any formula. 

Next consider the connective \emph{honk}: 

\begin{center}
 \AxiomC{$- \ A$}
\AxiomC{$+ \ B$}
\LeftLabel{$+\mathit{honk}I$:}
\BinaryInfC{$+ \ A\mathit{honk} B$}
\DisplayProof\qquad
\AxiomC{$+ \ A\mathit{honk} B$}
\LeftLabel{$+\mathit{honk} E$:}
\UnaryInfC{$- \ A$}
\DisplayProof\quad 
\AxiomC{$+ \ A\mathit{honk} B$}
\UnaryInfC{$+ \ B$}
\DisplayProof

\bigskip

\AxiomC{$+ \ A$}
\AxiomC{$- \ B$}
\LeftLabel{$- \mathit{honk} I$:}
\BinaryInfC{$- \ A \mathit{honk} B$}
\DisplayProof\qquad
\AxiomC{$- \ A\mathit{honk} B$}
\LeftLabel{$-\mathit{honk} E$:}
\UnaryInfC{$+ \ A$}
\DisplayProof\quad 
\AxiomC{$- \ A\mathit{honk} B$}
\UnaryInfC{$- \ B$}
\DisplayProof
\end{center}

\noindent If $conk$ is present in the logic, then, given the co-ordination principle, the assertion of any formula follows from the denial of any formula, and the denial of any formula follows from the assertion of any formula. 

The assertive and rejective rules of $conk$ and $honk$ do not matched up as demanded by conversion. Conversion thus solves a problem with bilateral versions of Prior's $tonk$.\footnote{$conk$ and $honk$ are also discussed in \cite{kurbisnormalbilateral}, which also contains the evident deductions showing their disastrous consequences.}

\section{Conclusion} 
In this paper I have given an account of bilateral harmony that is an alternative to the one proposed by Francez. The process of inversion described in section 4 corresponds to Francez's notion of vertical harmony when adapted to bilateral logic. The process of conversion described in section 5 corresponds to his notion of horizontal harmony. Inversion captures the conceptual considerations that the grounds and consequences of asserting formulas should exhibit a certain balance, and similarly for denying it. Conversion captures the conceptual considerations advanced in section 3 that the consequences of asserting a formula also give the grounds for denying it, namely if some of those consequences are denied, and that denying a formula gives the grounds for denying that some of the grounds for its assertion obtain, and analogously for the grounds and consequences of denying a formula. I proposed that every connective of bilateral logic be governed by rules of both kinds of the unilateral account of harmony presented in section 4. Thus each connective has verificationist and pragmatist aspects to its meaning. I imposed restrictions on the rules of inference of bilateral logic to streamline the exposition of the process of conversion. A more general account is possible and formulating it could be done in further work. I cordially invite the interested reader to join in this project and to contribute to improving the present account.

\bibliographystyle{eptcs}
\bibliography{biblio}

\end{document}